\documentclass{article}

\usepackage{arxiv}

\usepackage[utf8]{inputenc}
\usepackage[T1]{fontenc}

\usepackage{amsmath, amssymb, amsfonts, bm}

\usepackage{physics}       
\usepackage{siunitx}       

\usepackage{graphicx}
\usepackage{booktabs}
\usepackage{multirow}
\usepackage{array}
\newcolumntype{P}[1]{>{\centering\arraybackslash}p{#1}}

\usepackage{float}

\usepackage{xcolor}
\definecolor{rev}{rgb}{0,0,0}
\definecolor{rev2}{rgb}{0,0,0}

\usepackage{caption}
\usepackage{subcaption}    

\usepackage{enumitem}
\setlist[itemize]{leftmargin=*}
\setlist[enumerate]{leftmargin=*}

\usepackage{algorithm}
\usepackage{algpseudocode}

\usepackage{hyperref}
\hypersetup{
  colorlinks=true,
  linkcolor=blue,
  citecolor=blue,
  urlcolor=blue
}

\usepackage[capitalise]{cleveref}

\usepackage{nicefrac}      
\usepackage{microtype}     
\usepackage{relsize}
\usepackage{cancel}        

\newcommand{\dataset}[1]{\textsc{#1}}
\newcommand{\burgers}{\dataset{Burgers}}

\newcommand{\nskt}{\dataset{NSKT}}
\newcommand{\roadenkf}{\textsc{ROAD--EnKF}}

\usepackage{cite}

\title{Superresolving Non-linear PDE Dynamics with Reduced-Order Autodifferentiable Ensemble Kalman Filtering For Turbulence Modeling and Flow Regulation}

\author{
  Mrigank Dhingra \\
  Department of Mechanical and Aerospace Engineering,\\
  University of Tennessee,\\
  Knoxville, TN 37996, USA.\\
  \texttt{mdhingra@vols.utk.edu}
  \And
  Omer San \\
  Department of Mechanical and Aerospace Engineering,\\
  University of Tennessee,\\
  Knoxville, TN 37996, USA.\\
  \texttt{osan@utk.edu} 
}

\begin{document}

\maketitle

\begin{abstract}
Accurately reconstructing and forecasting high-resolution (HR) states from computationally cheap low-resolution (LR) observations is central to estimation-and-control of spatio-temporal PDE systems. We develop a unified superresolution pipeline based on the reduced-order autodifferentiable Ensemble Kalman filter (ROAD–EnKF). The method learns (i) a low-dimensional latent dynamics model and (ii) a nonlinear decoder from latent variables to HR fields; the learned pair is embedded in an EnKF, enabling simultaneous state estimation and control-oriented forecasting with quantified uncertainty. We evaluate on three benchmarks: 1-D viscous Burgers equation (shock formation), Kuramoto–Sivashinsky (KS) equation (chaotic dynamics), and 2-D Navier–Stokes–Kraichnan turbulence (NSKT) (vortex decaying dynamics at $\mathrm{Re}=16{,}000$). LR data are obtained by factors of $4$–$8$ downsampling per spatial dimension and are corrupted with noise. On Burgers and KS, the latent models remain stable far beyond the observation window, accurately predicting shock propagation and chaotic attractor statistics up to $150$ steps. On 2-D NSKT, the approach preserves the kinetic-energy spectrum and enstrophy budget of the HR data, indicating suitability for control scenarios that depend on fine-scale flow features. These results position ROAD–EnKF as a principled and efficient framework for physics-constrained superresolution, bridging LR sensing and HR actuation across diverse PDE regimes.

\end{abstract}

\section{\label{sec:level1}Introduction} 

\subsection{\label{sec:level2}Motivation}


Reconstructing and forecasting high–dimensional physical states from sparse, noisy measurements is central to geoscience, fluid dynamics, and control. Ensemble Kalman methods (EnKFs) address this by combining a model–based forecast with a Bayesian‐style analysis, but their cost and sensitivity to model error limit use at high spatial resolutions. A promising direction is to \emph{learn} reduced surrogates and use them \emph{inside} the filter.


A recent step in this direction is the \emph{Reduced–Order Autodifferentiable EnKF (ROAD–EnKF)} of Chen, Sanz–Alonso, and Willett~\cite{Chen_2023}. ROAD–EnKF learns (i) a low–dimensional latent dynamics
\[
z_t \;=\; G_{\alpha}(z_{t-1}) + \zeta_t,
\]
and (ii) a decoder mapping latent states back to the physical state,
\[
u_t \;=\; D_{\gamma}(z_t).
\]
Training maximizes an EnKF–estimated log–likelihood by differentiating through the EnKF, and the learned latent state–space model (SSM) is then used inside the EnKF for reconstruction and forecasting. ROAD–EnKF shows that when the dynamics admit a hidden low–dimensional structure one can obtain accuracy comparable to full–order approaches at \emph{much lower cost} and handle \emph{partial/noisy} observations.

We specialize the ROAD–EnKF paradigm to the \emph{low–resolution to high–resolution (LR\,$\to$\,HR) super–resolution} regime that arises when sensors deliver coarse fields but decision–making or downstream models require high resolution. Concretely, we assume we observe low–resolution fields \(y_t^{\mathrm{LR}}\) related to high–resolution states \(u_t^{\mathrm{HR}}\) by a known or learnable downsampling operator \(\mathcal{D}(\cdot)\):
\[
y_t^{\mathrm{LR}} \;=\; \mathcal{D}(u_t^{\mathrm{HR}}) + \eta_t.
\]
We learn the same latent SSM as in ROAD–EnKF, but we \emph{explicitly} design the decoder and the observation path for super–resolution and we train with SR–specific objectives. At test time, SRDA filters LR data to produce HR reconstructions with uncertainty and then forecasts in HR.

\subsection{Literature Review}

Traditional surrogates approximate the \emph{solution} of a parametric
PDE on a fixed grid, whereas \emph{neural operators} learn the entire
\emph{solution map}.
The Fourier Neural Operator (FNO) of
Li~\textit{et al.}~\cite{li2021fourierneuraloperatorparametric}
achieves mesh-independent generalisation by learning spectral
integral kernels, while DeepONet~\cite{lu2021deeponet}
provides a universal operator-approximation theorem via a
branch–trunk architecture.
Recent efforts push capacity and geometric flexibility:
HAMLET couples graph attention with transformers for irregular
meshes~\cite{bryutkin2024hamletgraphtransformerneural};
physics-informed transformer operators extend to general
IBVPs~\cite{boya2025physicsinformedtransformerneuraloperator};
Li~\textit{et al.} propose a geometry-informed neural operator for
large-scale 3-D domains~\cite{li2023geometryinformedneuraloperatorlargescale}; and
state-space models such as the Mamba Neural Operator offer
memory-efficient sequence processing~\cite{cheng2025mambaneuraloperatorwins}.
Kossaifi~\textit{et al.} introduce a multi-grid tensorised FNO that
compresses spectral weights via Tucker factorisation
\cite{kossaifi2023multigridtensorizedfourierneural}.
Complementary work learns data-driven discretizations
\cite{Bar_Sinai_2019} or embeds symmetries such as Galilean invariance.
Our decoder adopts the spectral-convolution paradigm but applies
frequency features only once, keeping inference costs low.

Compute efficiency has become a parallel focus:  
the \emph{decomposed FNO} (D-FNO) replaces costly 3-D FFTs with a
rank-decomposed series of 1-D FFTs, cutting the
$\mathcal O(N^{3}\log N)$ complexity to
$\mathcal O(PN\log N)$ while retaining accuracy
\cite{LI2025117732}.  
Physics-augmented latent compression further reduces memory via
low–dimensional latent grids \cite{Lu2024PhysicsAugmented},  
and a reduced–geostatistical FNO accelerates large 3-D inverse problems
in hydrogeology \cite{2024WRR....6034939G}.  
Parameter–efficient adaptation strategies mirror LoRA in NLP:
FouRA applies low-rank updates to spectral blocks
\cite{FouRA2024}, while
iFNO introduces an \emph{invertible} FNO that shares weights between
forward and inverse operators \cite{iFNO2024}.  
Collectively, these advances show that high-resolution fidelity and
computational tractability need not be mutually exclusive.

Learning to reconstruct high-resolution (HR) flow fields from coarse
data has gained traction as DNS remains prohibitive at realistic
Reynolds numbers.
Kim~\textit{et al.} employed GANs with physics-informed losses for
3-D turbulence super-resolution \cite{BODE20212617}.
Xie~\textit{et al.} stabilised temporal roll-outs via PDE residual
penalties~\cite{REN2023112438}.
For 2-D Kraichnan turbulence, Maulik and San showed that
data-driven deconvolution restores power-law spectra more faithfully
than classical filters~\cite{10.1063/1.5079582}.
Recent variants integrate recurrent memory and explicit constraints:
Wang~\textit{et al.}’s PIRGAN reconstructs rotating-detonation
combustor flows~\cite{WANG2024105649}, while
Khademi~\textit{et al.} combine domain discretization with
Sobolev-norm losses in DG-PINN for 3-D turbulence
\cite{KHADEMI2025104988}.
Our study differs by coupling a latent neural-operator surrogate with a
differentiable EnKF so that uncertainty is propagated and corrected
online.

Beyond GAN-based approaches, physics-augmented latent operators now
deliver SR reconstructions with wall-clock gains that scale favourably
with grid size \cite{Lu2024PhysicsAugmented}.

Snapshot Proper Orthogonal Decomposition (POD) remains the workhorse of
projection-based ROMs \cite{Sirovich1987_POD}.
Comprehensive surveys by Benner~\textit{et al.}
\cite{doi:10.1137/130932715} and Taira~\textit{et al.}
\cite{Taira2020_ModalROM} review balanced truncation, dynamic-mode
decomposition, and closure models.
Recent ML-based ROMs seek richer latent spaces:
$\beta$-VAEs with transformers capture chaotic
dynamics~\cite{solerarico2023betavariationalautoencoderstransformersreducedorder},
and attention-augmented autoencoders improve feature capture in
unsteady flows~\cite{BEIKI2025110463}.
Our latent space serves as a \emph{learned}, nonlinear ROM trained
jointly with a dynamics model (Sect.~\ref{sec:latent_dynamics}). 

Multi-level PINN hierarchies also lower solve times for large
structural-mechanics PDEs by recursively coarsening the residual loss
\cite{He2024MultiLevel}.

Foundational work by Lions~\cite{Lions1971_OptCtrlPDE} established
functional-analytic conditions for existence of optimal controls.
Adjoint-based algorithms were formalised in the monographs of
Hinze~\textit{et al.} \cite{Hinze2008_PDEOpt} and
Tröltzsch \cite{Troeltzsch2010_OC_PDE}.
Recent ML-oriented approaches include adjoint-oriented neural networks
(AONN) for all-at-once optimization \cite{yin2023aonnadjointorientedneuralnetwork}
and PINN frameworks for PDE-constrained control
\cite{MOWLAVI2023111731}.
Although we do not solve a classical open-loop problem, our latent
dynamics are trained \emph{jointly} with the EnKF loss, effectively
learning a feedback control that steers forecasts towards observations.

The Ensemble Kalman Filter (EnKF) balances Gaussian assumptions and
particle degeneracy; see Navon’s review \cite{Navon2005_DAComp}.
Reduced-Order Autodifferentiable EnKF (ROAD-EnKF) enables back-prop
through the Kalman gain~\cite{Chen_2023}.
Hybrid schemes now combine deep surrogates with classical DA:
Chattopadhyay~\textit{et al.} augment EnKF with data-driven
ensembles~\cite{CHATTOPADHYAY2023111918};
Peng~\textit{et al.} fuse DL and DA for model-error estimation
\cite{Peng2024Hybrid};
and Patel~\textit{et al.} augment PINNs with turbulence models for
mean-flow reconstruction \cite{PhysRevFluids.9.034605}.
Our framework inherits ROAD-EnKF’s differentiability and extends it to
super-resolution, allowing latent corrections while the decoder bridges
scales.

A complementary Bayesian perspective places priors on operator weights:
VB-DeepONet performs variational inference for operator learning
\cite{GARG2023105685},  
Approximate Bayesian Neural Operators propagate parameter uncertainty
through predictions \cite{magnani2022approximatebayesianneuraloperators},  
and ProbNO extends the idea to fully probabilistic functionals
\cite{ProbNO2025}.  
Such methods highlight the growing demand for principled uncertainty
quantification—addressed here via ensemble spread and, in future work,
potentially via Bayesian neural operators.

Brunton, Noack and Koumoutsakos chart the rise of ML in fluid
mechanics \cite{Brunton2020_MLFluid}.
Duraisamy~\textit{et al.} emphasise uncertainty quantification when
replacing subgrid closures with data-driven surrogates
\cite{Duraisamy2019_TurbData}.  
Our ensemble-based formulation provides both HR reconstructions and
uncertainty bounds, answering this call.

Differentiable-simulation toolkits are making gradient-based design
routine; Newbury~\textit{et al.} survey over a dozen such frameworks
\cite{newbury2024reviewdifferentiablesimulators}, and
Joglekar introduces \emph{generative neural re-parameterisation} to
amortise PDE-constrained optimization across control queries
\cite{joglekar2024generativeneuralreparameterizationdifferentiable}.
These trends motivate our use of an end-to-end differentiable filter.

Despite impressive individual advances, few works integrate neural
operators, super-resolution, and differentiable EnKF into a single control based
pipeline.  
By coupling a low-dimensional latent neural operator with ROAD-EnKF we
achieve (i) end-to-end learning of dynamics and observation models,
(ii) physically consistent HR reconstructions with quantified
uncertainty, and (iii) computational efficiency suitable for
optimal-control loops.  
The following sections demonstrate these benefits on Burgers, KS and
NSKT benchmarks.

\begin{figure*}[ht]
\includegraphics[width=\textwidth]{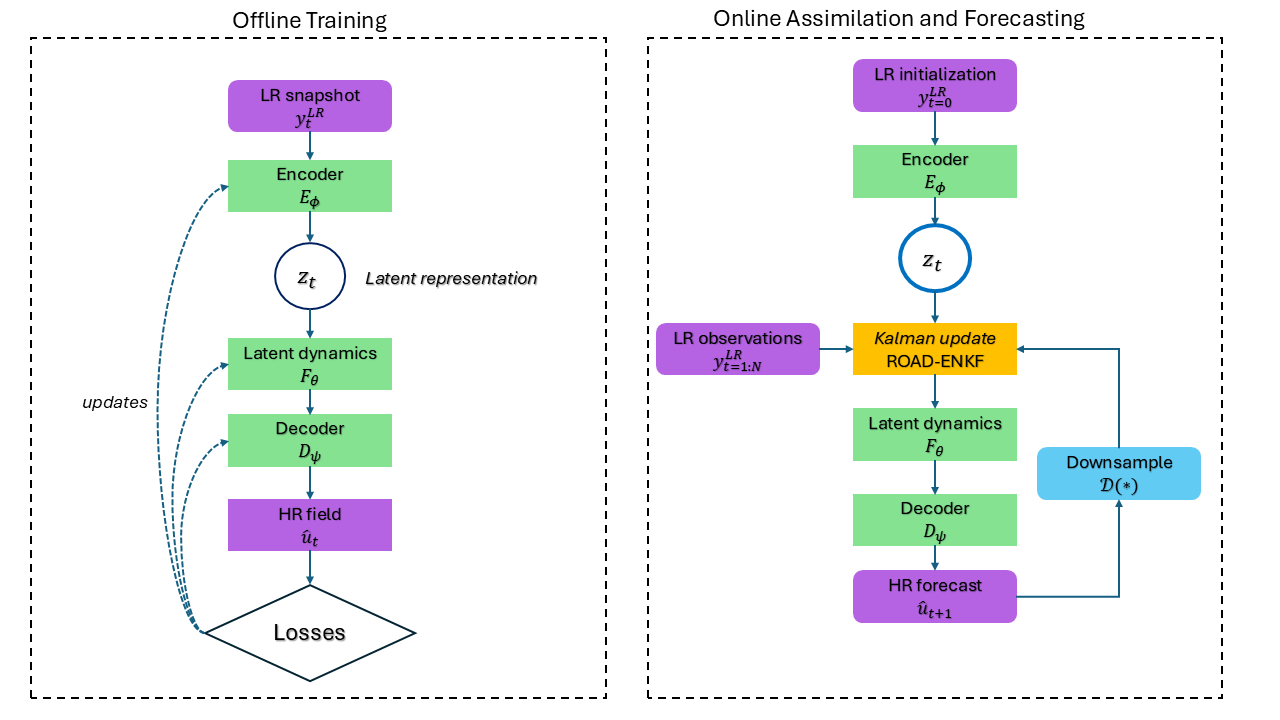}
\caption{\label{fig:SRDA_illust}Conceptual workflow of the \emph{super-resolution ROAD-EnKF} framework. 
\textbf{Left panel – Offline training.} A low-resolution (LR) snapshot $y_t^{\text{LR}}$
is embedded by the encoder $E_{\phi}$ into a latent representation $z_t$, advanced
one step by the latent dynamics operator $F_{\theta}$ and decoded by
$D_{\psi}$ to a high-resolution (HR) field $\hat u_t$. The network parameters
$\{\phi,\theta,\psi\}$ are updated through back-propagation of a composite loss
comprising an LR reconstruction term, an HR supervision term and a latent
regularisation penalty. 
\textbf{Right panel – Online assimilation and forecasting.}
At $t=0$ the latent ensemble is initialised from an LR state $y_0^{\text{LR}}$.
For each subsequent time step, the predicted latent ensemble is corrected by a
\emph{Kalman update} (ROAD-EnKF) that assimilates the current LR observation
$y_t^{\text{LR}}$ after passing it through the same decoder–down-sample–encoder
measurement operator used during training. The updated latent state is
propagated with $F_{\theta}$ and decoded to produce the HR forecast
$\hat u_{t+1}$; this forecast is then recycled inside the loop to continue the
assimilation–forecast cycle.}
\end{figure*}

\subsection{Reduced-Order Autodifferentiable EnKF} \label{sec:road-enkf}

The Reduced--Order Autodifferentiable Ensemble Kalman Filter is a hybrid learning--filtering framework that couples a learnt latent surrogate with an
online Ensemble Kalman Filter (EnKF) \cite{Chen_2023}.  By performing data assimilation in a
low-dimensional latent subspace, ROAD--EnKF (i) circumvents the cubic cost of
full-state EnKF updates and (ii) allows gradients to flow through the complete
forecast--analysis cycle, enabling end-to-end optimization with stochastic
gradient descent.

Let $E_{\phi}$, $F_{\theta}$ and $D_{\psi}$ denote the LR encoder, latent
dynamics, and HR decoder, respectively.  For an LR trajectory
$\{\mathbf y_t^{\mathrm{LR}}\}_{t=0}^{T-1}$, the filter propagates an ensemble
$\{\mathbf z_{t,i}\}_{i=1}^N$ in latent space while repeatedly reconciling
model forecasts with new observations.  The complete algorithm for one
mini-batch is summarised in Alg.~\ref{alg:road_enkf}.

\begin{algorithm}[H]
\caption{ROAD--EnKF forecast--analysis cycle (single mini-batch)}
\label{alg:road_enkf}
\begin{algorithmic}[1]
  \Require LR sequence $\{\mathbf y_t^{\mathrm{LR}}\}_{t=0}^{T-1}$, ensemble size $N$, prior std.\ $\sigma_{\text{prior}}$, update period $s_{\text{upd}}$, noise cov.\ $R$
  \State \textbf{Encode:} $\mathbf z_0 \gets E_{\phi}(\mathbf y^{\mathrm{LR}}_0)$
  \State \textbf{Initial ensemble:} $Z_0 \gets \mathbf z_0 \mathbf 1_N^{\mathsf T} + \sigma_{\text{prior}}\,\varepsilon$, \quad $\varepsilon \sim \mathcal N(0,\sigma^2 I)$
  \For{$t \gets 0$ \textbf{to} $T-1$}
    \If{$t \bmod s_{\text{upd}} = 0$}
      \State Decode $Z_t$ to HR, down-sample $\rightarrow Y_t^{\mathrm f}$
      \State Compute anomalies $A_z$, $A_y$ and Kalman gain $K$ \Comment{see Sec.~\ref{sec:enkf_update}}
      \State Draw perturbed obs.\ $\tilde{\mathbf y}_t \sim \mathcal N(\mathbf y_t^{\mathrm{LR}}, R)$
      \State $Z_t \gets Z_t + K\bigl(\tilde{\mathbf y}_t - Y_t^{\mathrm f}\bigr)$
    \EndIf
    \If{$t = T-1$}
      \State \textbf{break}
    \EndIf
    \State \textbf{Prediction:} $Z_{t+1} \gets F_{\theta}(Z_t)$
  \EndFor
  \State \Return final ensemble $Z_{T-1}$
\end{algorithmic}
\end{algorithm}

During training the stochastic residuals introduced by the perturbed
observations propagate gradients back to the encoder, latent dynamics, and
decoder, letting the network learn to minimise the joint forecast error over
full trajectories while respecting the Bayesian update structure of the
EnKF.

\section{Datasets}

To benchmark the proposed SR-enabled ROAD-EnKF pipeline rigorously, we assemble three complementary datasets spanning synthetic turbulence, canonical geophysical flows, and real-world observations.  

\subsection{Viscous \burgers{} Equation}\label{subsec:burgers}
\begin{figure}
    \centering
    \begin{subfigure}[b]{0.45\textwidth}
        \centering
        \includegraphics[width=\textwidth]{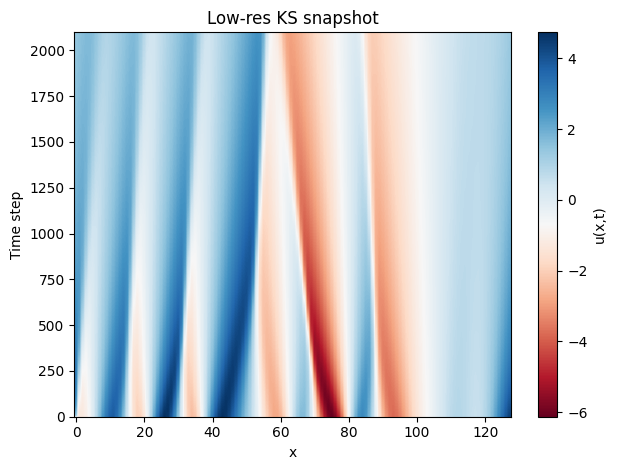}
        \caption{Low-resolution Burgers 1D data}
        \label{fig:lr_burgers}
    \end{subfigure}
    \hfill
    \begin{subfigure}[b]{0.45\textwidth}
        \centering
        \includegraphics[width=\textwidth]{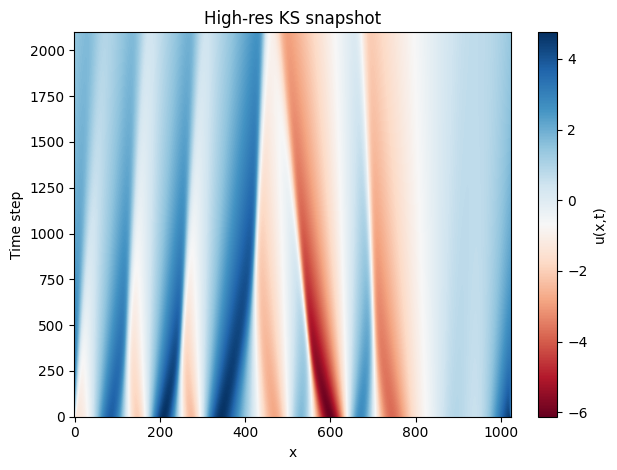}
        \caption{High-resolution Burgers 1D data}
        \label{fig:hr_burgers}
    \end{subfigure}
    \caption{LR and HR Burgers 1D datasets for 2000 timesteps. The LR data has been downscaled by a factor of 8.}
    \label{fig:burgers_lr_hr}
\end{figure}

We consider the dimensionless viscous 1‑D \burgers{} equation (see Fig. \ref{fig:burgers_lr_hr}) on the unit circle,
\begin{equation}
\partial_t u(x,t) + u(x,t)\,\partial_x u(x,t) = \nu\,\partial_{xx}u(x,t), \qquad x\in[0,1),\; t>0,\label{eq:burgers}
\end{equation}
subject to periodic boundary conditions $u(0,t) = u(1,t)$ and an initial condition $u(x,0)=u_0(x)$ \cite{burgers1948}.  Here $u$ denotes the scalar velocity field and $\nu$ is the kinematic viscosity; throughout this work we fix $\nu=10^{-2}$.

The spatial domain is discretized into $N_{\mathrm{HR}} = 1024$ equispaced points with grid spacing $\Delta x = 1/N_{\mathrm{HR}}$.  Central finite differences approximate the first and second derivatives:
\begin{align}
\partial_x u &\approx \frac{u_{j+1}-u_{j-1}}{2\,\Delta x}, &
\partial_{xx} u &\approx \frac{u_{j+1}-2u_j+u_{j-1}}{\Delta x^{2}},\label{eq:fd}
\end{align}
which, when substituted into~\eqref{eq:burgers}, yields the semi‑discrete right-hand side
\begin{equation}
\mathrm{RHS}_j = -u_j\,\frac{u_{j+1}-u_{j-1}}{2\,\Delta x} + \nu\,\frac{u_{j+1}-2u_j+u_{j-1}}{\Delta x^{2}}.\label{eq:rhs}
\end{equation}
Time integration is performed with an explicit fourth‑order Runge–Kutta (RK4) scheme.  The CFL‑stable step size is chosen adaptively as
\begin{equation}
\Delta t = \mathrm{CFL}\times \min\!\bigl( \tfrac{\Delta x}{\max_j |u_j| + 10^{-12}},\; \tfrac{\Delta x^{2}}{2\nu}\bigr), \qquad \mathrm{CFL}=0.4.
\end{equation}

The initial velocity field is synthesized as a random superposition of ten sine modes,
\begin{equation}
 u_0(x) = \sum_{i=1}^{10} \sin\bigl(2\pi k_i x + \phi_i\bigr), \qquad k_i\sim\mathcal U(1,10),\; \phi_i\sim\mathcal U(0,2\pi),\label{eq:u0}
\end{equation}
where $\mathcal U$ denotes the continuous uniform distribution.  The RK4 solver is run until $T_{\text{end}}=0.04$, producing a trajectory $
{\sim}2000$ snapshots of the HR field $u^{\mathrm{HR}}(t)\in\mathbb R^{1024}$.  Low‑resolution data are obtained by uniform subsampling with factor $s=8$:
\begin{equation}
 u^{\mathrm{LR}}_j(t) = u^{\mathrm{HR}}_{8j}(t), \qquad j = 0,\dots,127.
\end{equation}
The resulting dataset dimensionalities are
\begin{align*}
\text{HR:}\ &\;(N_t, N_{\mathrm{HR}}) = (2001,1024), &
\text{LR:}\ &\;(N_t, N_{\mathrm{LR}}) = (2001,128).
\end{align*}
LR snapshots are further corrupted with additive Gaussian noise at signal‑to‑noise ratio of \SI{30}{dB} before being assimilated by the \roadenkf{} filter.

\subsection{Two‐Dimensional Navier–Stokes–Kraichnan Turbulence}\label{subsec:NSKT}

\begin{figure}
    \centering
    \includegraphics[width=0.75\linewidth]{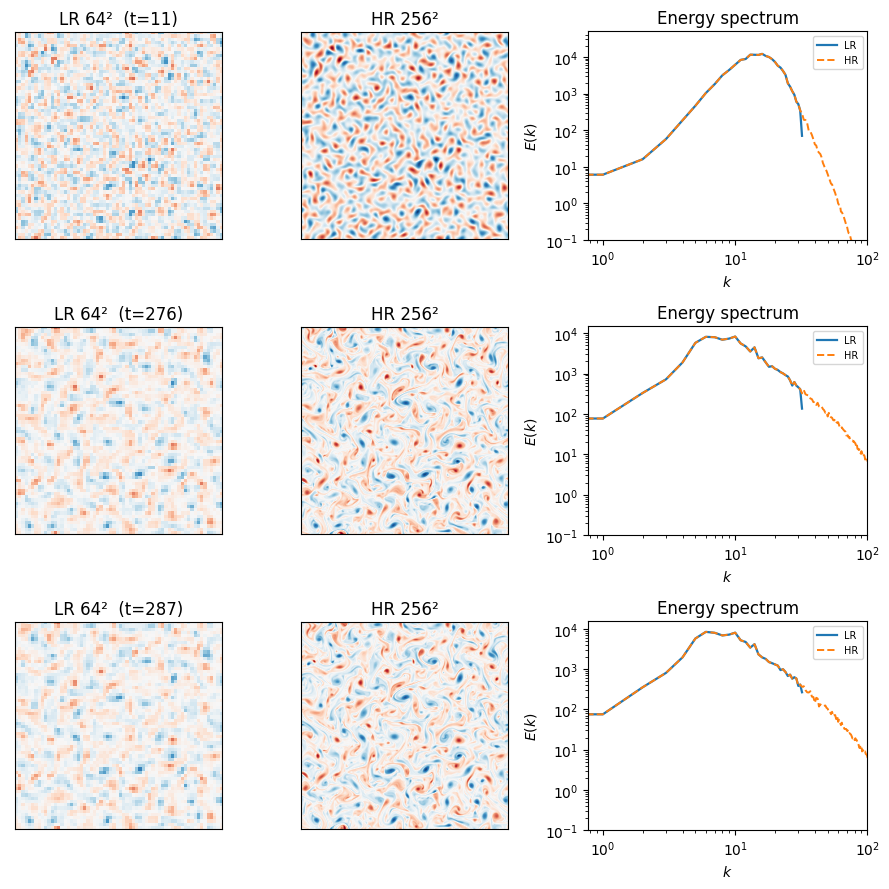}
    \caption{Low resolution (64x64) and high resolution (256x256) NSKT dataset. Bicubic downsampling was used to generate the low resolution data.}
    \label{fig:nskt_lr_hr}
\end{figure}

The evolution of the vorticity field $\omega$ in 2D turbulence (see Fig. \ref{fig:nskt_lr_hr}) is governed by the incompressible Navier-Stokes equation in the vorticity-streamfunction formulation \cite{ErichsonSuperBench}:
\begin{equation}
    \frac{\partial \omega}{\partial t} + \bm{u} \cdot \nabla \omega = \nu \nabla^2 \omega + f(\bm{x}, t),
    \label{eq:navier_stokes_vorticity}
\end{equation}
where:
\begin{itemize}
    \item $\omega = \nabla \times \bm{u}$ is the vorticity field, with $\bm{u}$ being the velocity vector field.
    \item $\nu$ is the kinematic viscosity, controlling the dissipation of small-scale structures.
    \item $f(\bm{x}, t)$ is an external forcing term, injecting energy into the system at specific scales.
    \item $\bm{u}$ is related to the streamfunction $\psi$ via:
    \begin{equation}
        \bm{u} = \nabla^\perp \psi = \begin{pmatrix} -\frac{\partial \psi}{\partial y} \\ \frac{\partial \psi}{\partial x} \end{pmatrix},
    \end{equation}
    where $\nabla^2 \psi = -\omega$ provides the relationship between the vorticity and streamfunction.
\end{itemize}

The energy spectrum $E(k)$ quantifies the distribution of kinetic energy across wavenumbers $k$. In Kraichnan's 2D turbulence, the energy cascade results in:
\begin{itemize}
    \item \textbf{Inverse Cascade:} Energy is transferred to larger scales (lower $k$), leading to the formation of coherent structures.
    \item \textbf{Direct Cascade:} Enstrophy (mean-square vorticity) is transferred to smaller scales (higher $k$), eventually dissipated by viscosity.
\end{itemize}

Forcing is applied in the wavenumber range $k_f$, typically targeting intermediate scales to sustain turbulence. The energy input rate $\epsilon$ is balanced by dissipation, ensuring statistical stationarity.

The dataset used in this study was generated by numerically solving the vorticity equation \eqref{eq:navier_stokes_vorticity} with the following parameters:
\begin{itemize}
    \item \textbf{Reynolds Number:} $\text{Re} = 16,000$, ensuring a highly turbulent regime.
    \item \textbf{Grid Resolution:} $256 \times 256$, providing sufficient resolution to capture multiscale dynamics.
    \item \textbf{Time Stepping:} A small time step of $\Delta t = 10^{-4}$ was used for numerical stability and accuracy.
    \item \textbf{Total Simulation Time:} $T = 8$ seconds, corresponding to 80,000 timesteps.
    \item \textbf{Forcing Wavenumber Range:} $k_f \in [4, 6]$, injecting energy at intermediate scales.
    \item \textbf{Boundary Conditions:} Periodic in both spatial directions.
\end{itemize}

The vorticity fields were saved at 800 equispaced timesteps, resulting in a dataset of shape $(800, 3, 256, 256)$, where the second dimension corresponds to velocity components $u$, $v$, and vorticity $\omega$. 

Low‐resolution (LR) counterparts $\omega^{\text{LR}}_{n}\in\mathbb R^{64\times64}$ are generated via bicubic interpolation preceded by an ideal low‐pass anti‐aliasing filter that removes wavenumbers $|\mathbf k|>k_c=\tfrac13 N_x$.

\subsection{Kuramoto–Sivashinsky Dataset}
\label{sec:data:ks}
\begin{figure}
    \centering
    \begin{subfigure}[b]{0.45\textwidth}
        \centering
        \includegraphics[width=\textwidth]{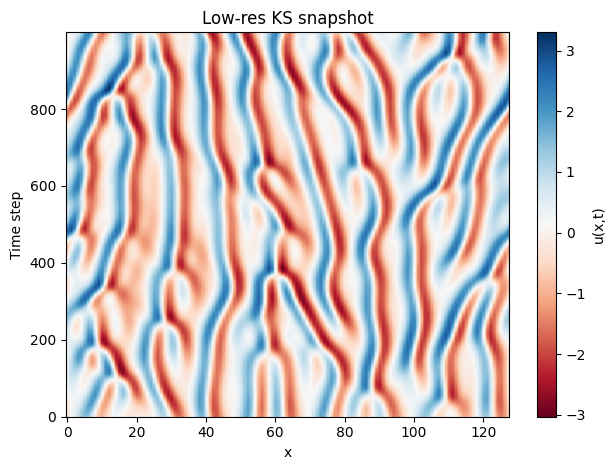}
        \caption{Low-resolution KS data}
        \label{fig:lr_ks}
    \end{subfigure}
    \hfill
    \begin{subfigure}[b]{0.45\textwidth}
        \centering
        \includegraphics[width=\textwidth]{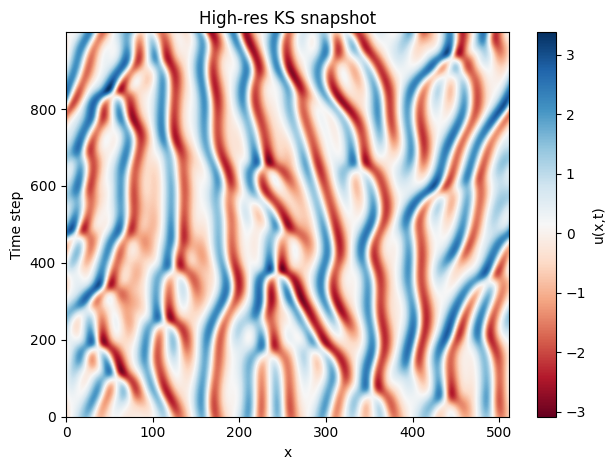}
        \caption{High-resolution KS data}
        \label{fig:hr_ks}
    \end{subfigure}
    \caption{LR and HR KS datasets for 1000 timesteps. The LR data has been downscaled by a factor of 4.}
    \label{fig:ks_lr_hr}
\end{figure}

The Kuramoto–Sivashinsky (KS) equation captures chaotic dynamics (see Fig. \ref{fig:ks_lr_hr}) of flame fronts and thin liquid films \cite{phlippe2024kuramoto}. In one spatial dimension with periodic boundary conditions it reads
\begin{equation}
  \partial_t u + u\,\partial_x u + \partial_{xx}u + \partial_{xxxx}u = 0, \qquad x \in [0, L],\; t>0.
  \label{eq:ks}
\end{equation}
Here $L = 32\pi$ is chosen to ensure fully developed spatio‑temporal chaos.

 We employ the exponential time‑differencing fourth‑order Runge–Kutta (ETDRK4) algorithm. Let $k$ be the Fourier wavenumbers on an equispaced grid with $N_{\mathrm{hr}}=512$ points; then writing $u(x,t) = \sum\hat{u}_k(t)e^{ikx}$, equation~\eqref{eq:ks} becomes
\begin{equation}
  \partial_t \hat{u}_k = -(k^{2}+k^{4})\,\hat{u}_k - \frac{i k}{2}\;\widehat{u^{2}}_{k},
\end{equation}
where the nonlinear term is evaluated pseudo‑spectrally. Pre‑computed ETDRK4 coefficients allow stable advancement with sub‑step $\Delta t = 10^{-5}$.

 Starting from small‑amplitude Gaussian noise, we discard an initial \num{2\,000\,000} sub‑steps (\SI{20}{\text{physical time units}}) as burn‑in to reach the chaotic attractor. We then integrate for an additional \SI{100}{} time units, recording a snapshot every $0.1$ units (every \num{10\,000} sub‑steps), yielding $T=1000$ high‑resolution states $u^{(t)} \in \mathbb{R}^{512}$.

 For data assimilation we construct coarse measurements by average‑pooling contiguous blocks of size four, resulting in $N_{\mathrm{lr}}=128$‑point signals
\begin{equation}
  u^{(t)}_{\mathrm{lr}}[j] \,=\, \frac{1}{4}\sum_{m=0}^{3} u^{(t)}_{\mathrm{hr}}[4j+m], \qquad j = 0,\dots,127.
\end{equation}
This mimics a sensor array with limited spatial fidelity while preserving large‑scale statistics.

\section{Methodology}

The methodology section opens with a high-level roadmap to guide readers through the pipeline.  We first formalize the SR observation operator that maps coarse states onto the latent high-resolution grid.  Next, we review the ensemble Kalman filter (EnKF) update and describe the spectral-space localization used to curb spurious long-range correlations.  Finally, we detail the training strategy for the neural SR prior.

\subsection{Latent Dynamics Model} \label{sec:latent_dynamics}
State estimation is performed in a learned latent space of dimension $d_z$ (typically $d_z\ll N_{\mathrm{LR}}$) governed by a stochastic residual Euler discretization
\begin{equation}\label{eq:latent_evol}
\bm{z}_{t+1} = \bm{z}_t + \Delta t\,f_{\boldsymbol{\theta}}(\bm{z}_t) + \boldsymbol{\varepsilon}_t,\qquad \boldsymbol{\varepsilon}_t \sim \mathcal{N}\bigl(\bm{0},\sigma^2\mathbf{I}\bigr),
\end{equation}
where $f_{\boldsymbol{\theta}}\!:\mathbb{R}^{d_z}\!\to\mathbb{R}^{d_z}$ is a neural network parameterised by weights $\boldsymbol{\theta}$. Specifically, $f_{\boldsymbol{\theta}}$ is a two‑layer multilayer perceptron (MLP)
\begin{equation}
 f_{\boldsymbol{\theta}}(\bm{z}) = W_2\,\mathrm{ReLU}(W_1\bm{z}+\bm{b}_1)+\bm{b}_2, \qquad W_1\in\mathbb{R}^{h\times d_z},\;W_2\in\mathbb{R}^{d_z\times h},
\end{equation}
with hidden dimension $h$ and ReLU activation. The process noise scale $\sigma$ is either fixed \emph{a priori} or learned via back‑propagation by parameterising $\sigma=\exp(\alpha)$ with \mbox{$\alpha\in\mathbb{R}$}. 
Equation~\eqref{eq:latent_evol} is differentiable with respect to both $\bm{z}_t$ and $\boldsymbol{\theta}$, enabling seamless integration with the ROAD‑EnKF update and end‑to‑end optimization of the latent dynamics alongside the decoder and observation operators.

\subsection{Decoder Architecture} \label{sec:decoder}
The super–resolution decoder maps a latent state $\bm{z}_t\in\mathbb{R}^{d_z}$ returned by the latent dynamics model (Section~\ref{sec:latent_dynamics}) to a high–resolution physical field $\hat{u}_t \in\mathbb{R}^{N_{\mathrm{HR}}}$.  It is conceptually divided into three stages: (i) a seed fully–connected projection, (ii) a Fourier positional–encoding concatenation, and (iii) a convolutional refinement stack.

 A linear layer initialises a hidden feature map of width $C_h$:
\begin{equation}
  \label{eq:seed}
  \bm{h}_0 = W_{\mathrm{fc}}\,\bm{z}_t + \bm{b}_{\mathrm{fc}}\,\in\mathbb{R}^{C_h}.
\end{equation}
The vector $\bm{h}_0$ is broadcast along the spatial dimension to form a constant feature channel $\bm{H}_0(x)\in\mathbb{R}^{C_h}$ for every grid point $x\in[0,1]$.

Following we enrich each spatial location with $F$ sinusoidal embeddings
\begin{equation}
  \label{eq:fourier}
  \bm{\phi}(x) = \bigl[\sin(2\pi kx),\, \cos(2\pi kx)\bigr]_{k=1}^{F}\;\in\mathbb{R}^{2F}.
\end{equation}
Concatenating $\bm{\phi}(x)$ to $\bm{H}_0(x)$ yields the composite input tensor
\begin{equation}
  \bm{H}_{\mathrm{in}}(x) = \begin{bmatrix}\bm{H}_0(x)\\ \bm{\phi}(x)\end{bmatrix} \in \mathbb{R}^{C_h+2F}.
\end{equation}

A stack of $L$ one–dimensional convolutions (kernel size~3, circular padding) incrementally mixes local and global information:
\begin{equation}
  \bm{H}_{\ell+1}(x) = \sigma\!\bigl(\,\mathrm{Conv1D}_{\ell}\bigl(\bm{H}_{\ell}(\cdot)\bigr)(x)\bigr),\quad\ell=0,\dots,L-1,
\end{equation}
where $\sigma(\cdot)$ denotes the ReLU activation.  The final hidden map $\bm{H}_{L}(x)$ is projected to the physical field via a $1\times1$ convolution (point–wise linear map):
\begin{equation}
  \hat{u}_t(x) = \mathrm{Conv1D}_{\text{out}}\bigl(\bm{H}_{L}(\cdot)\bigr)(x)\;\in\mathbb{R}.
\end{equation}

Unless stated otherwise we set $d_z=64$, $C_h=128$, $F=32$, and $L=4$.  These values were selected to balance reconstruction accuracy and computational footprint.  The overall decoder contains roughly $\num{0.9}\,\text{M}$ trainable parameters and contributes less than \SI{5}{ms} of wall–clock latency per forward pass when evaluated on an NVIDIA RTX~4090 GPU.

Equation~\eqref{eq:seed} provides a global conditioning of the output on the latent code, while the sinusoidal basis~\eqref{eq:fourier} injects multi–scale positional context akin to classical Fourier spectral methods.  The convolutional stack then learns a residual mapping that blends these two sources of information, enabling the network to synthesise fine–scale structures consistent with both the learned dynamics and the observed low–resolution constraints.

\subsection{Latent Low Resolution Encoder} \label{sec:LRencode}
Given a flattened LR observation vector $\bm y_{\mathrm{LR}}\in\mathbb R^{N_{\mathrm{LR}}}$ we infer an initial latent state $\bm z_0$ through a simple two--layer multilayer perceptron (MLP)
\begin{align}
    \bm h &= \sigma\bigl(W_1\,\bm y_{\mathrm{LR}} + \bm b_1\bigr), && W_1\in\mathbb R^{128\times N_{\mathrm{LR}}}, \label{eq:enc-hidden}\\[4pt]
    \bm z_0 &= W_2\,\bm h + \bm b_2, && W_2\in\mathbb R^{d_z\times128},\label{eq:enc-out}
\end{align}
where $\sigma$ denotes the ReLU activation. Equations~\eqref{eq:enc-hidden}--\eqref{eq:enc-out} define the deterministic encoder map $g_{\phi}\colon\mathbb R^{N_{\mathrm{LR}}}\to\mathbb R^{d_z}$ with parameter set $\phi=\{W_1,W_2,\bm b_1,\bm b_2\}$. The modest hidden width of 128 suffices because spatial correlations are largely removed during LR downsampling; the network's main duty is dimensionality reduction and feature extraction.

For 2-D inputs we adopt a row-major flattening convention, i.e.\ $\bm y_{\mathrm{LR}} = \text{vec}\bigl(u_{\mathrm{LR}}(x_i,y_j)\bigr)$. Prior to encoding, LR snapshots are standardised to zero mean and unit variance across the training set.

\subsection{Ensemble Kalman Filter Assimilation} \label{sec:enkf_update}
Let
\(
  Z^{\mathrm f} = [\mathbf z^{\mathrm f}_1,\dots,\mathbf z^{\mathrm f}_N]
  \in\mathbb R^{d_z\times N}
\)
denote the \emph{forecast} ensemble in latent space, where \(d_z\) is the
latent dimension and \(N\) the ensemble size.  
Each particle is decoded to a high-resolution state and projected to the
sensor grid, yielding the forecast observation ensemble
\(
  Y^{\mathrm f} = [\mathbf y^{\mathrm f}_1,\dots,\mathbf y^{\mathrm f}_N]
  \in\mathbb R^{d_y\times N}
\),
with  
\(
  \mathbf y^{\mathrm f}_i
  =\mathcal D\!\bigl(\mathcal G(\mathbf z^{\mathrm f}_i)\bigr),
\)
where \(\mathcal G\) is the Fourier decoder
(Sect.~\ref{sec:decoder}) and \(\mathcal D\) the bicubic
down–sampling operator.

Define ensemble means
\(
  \bar{\mathbf z}=\tfrac1N\sum\nolimits_{i=1}^N\mathbf z^{\mathrm f}_i,
  \;
  \bar{\mathbf y}=\tfrac1N\sum\nolimits_{i=1}^N\mathbf y^{\mathrm f}_i
\),
and anomaly matrices
\[
  A_z = Z^{\mathrm f} - \bar{\mathbf z}\mathbf 1_N^{\mathsf T},
  \qquad
  A_y = Y^{\mathrm f} - \bar{\mathbf y}\mathbf 1_N^{\mathsf T}.
\]

With diagonal sensor noise covariance
\(R=\mathrm{diag}(\sigma_1^2,\dots,\sigma_{d_y}^2)\),
the standard EnKF gain
\(
  K = C_{zy}\,C_{yy}^{-1}
\)
is written in sample form as
\[
  K
  =\frac{A_zA_y^{\mathsf T}}{N-1}
    \Bigl(
      \frac{A_yA_y^{\mathsf T}}{N-1}+R
    \Bigr)^{\!-1}.
\]
Because \(d_y\gg N\), we avoid a large \(d_y\times d_y\) inverse by
recasting the problem in the ensemble space using the Woodbury identity:
\[
  G \;=\;
  \Bigl(
        I_N
      + A_y^{\mathsf T}R^{-1}A_y/(N-1)
  \Bigr)^{-1},
  \qquad
  K \;=\;
  \frac{A_zA_y^{\mathsf T}R^{-1}G}{N-1}.
\]
The \(N\times N\) matrix \(G\) is factorised via a Cholesky decomposition.

Perturbed observations are formed as
\(
  \tilde{\mathbf y}_i
  = \mathbf y^{\text{obs}} + \boldsymbol\varepsilon_i,
  \;
  \boldsymbol\varepsilon_i\sim\mathcal N(0,R).
\)
Each latent particle is then updated by
\[
  \mathbf z^{\mathrm a}_i
  \;=\;
  \mathbf z^{\mathrm f}_i
  + K\bigl(\tilde{\mathbf y}_i - \mathbf y^{\mathrm f}_i\bigr),
  \quad
  i=1,\dots,N.
\]
The \emph{analysis} ensemble
\(Z^{\mathrm a}\) serves as the initial condition for the
stochastic residual–Euler propagation described in
Sect.~\ref{sec:latent_dynamics}.

The overall complexity scales as
\(\mathcal O\!\bigl(N^3 + d_zN^2\bigr)\),
independent of the high-dimensional sensor space \(d_y\).
For our experiments with \(N\!=\!32\) and \(d_z\!\le\!64\),
the update step adds only milliseconds per assimilation window,
making it negligible compared with the decoder’s forward pass.

\subsection{Training Strategy and Hyperparameters} \label{sec:training}
The ROAD-EnKF network is trained end-to-end with a
\emph{sequential forecast–analysis loop} that mimics on-line data
assimilation.  Each mini-batch contains $B$ trajectories of length
$T_{\mathrm{seq}}$ low-resolution (LR) snapshots
$\{\mathbf y^{\mathrm{LR}}_{t}\}_{t=0}^{T_{\mathrm{seq}}-1}$
together with their high-resolution (HR) references
$\{\mathbf u^{\mathrm{HR}}_{t}\}_{t=0}^{T_{\mathrm{seq}}-1}$.

The LR encoder (Sect.~\ref{sec:LRencode}) maps the first observation to an
\emph{analysis} latent state
$
  \mathbf z_{0}
  = E_{\phi}\!\bigl(\mathbf y^{\mathrm{LR}}_{0}\bigr)
$.
An ensemble of $N$ particles is initialised as
$
  Z_{0}= \mathbf z_{0}\mathbf 1_{N}^{\mathsf T}
  +\sigma_{\text{prior}}\,
  \boldsymbol\eta,\;
  \boldsymbol\eta\sim\mathcal N(0,I),
$
and refined with a single EnKF update
(Alg.~\ref{sec:enkf_update}) against the first LR frame.

For every time index $t$:
\begin{enumerate}
  \item \textbf{Prediction.}  
        The latent ensemble is advanced through the residual-Euler
        model
        $
          Z^{\mathrm{f}}_{t+1}
          = \mathcal F_{\theta}\!\bigl(Z^{\mathrm{a}}_{t}\bigr)
        $
        (Sect.~\ref{sec:latent_dynamics}).
  \item \textbf{Assimilation.}  
        Every $s_{\text{upd}}\!=\!1$ step  
        the forecast ensemble is assimilated with the new LR
        observation to produce
        $
          Z^{\mathrm{a}}_{t+1}
        $.
        Between updates the filter runs purely in prediction mode.
  \item \textbf{Loss accumulation.}  
        The ensemble mean
        $
          \bar{\mathbf z}_{t+1} = \frac1N\sum_i\mathbf z^{\mathrm a}_{t+1,i}
        $
        is decoded to HR,
        $
          \hat{\mathbf u}_{t+1}=D_{\psi}(\bar{\mathbf z}_{t+1}),
        $
        and down-sampled,
        $
          \hat{\mathbf y}_{t+1}=\mathcal D(\hat{\mathbf u}_{t+1}).
        $
        We accumulate
        \[
          \mathcal L_t
          =
          \|\hat{\mathbf y}_{t+1}-\mathbf y^{\mathrm{LR}}_{t+1}\|_2^{2}
          +\lambda_{\mathrm{HR}}
           \|\hat{\mathbf u}_{t+1}-\mathbf u^{\mathrm{HR}}_{t+1}\|_2^{2}
          +\lambda_{\mathrm{lat}}
           \|\bar{\mathbf z}_{t+1}\|_2^{2}.
        \]
  \item \textbf{Truncated BPTT.}  
        After $\tau_{\text{TBPTT}}=\tfrac12T_{\mathrm{seq}}$ steps the
        latent gradient graph is detached to bound memory footprint.
\end{enumerate}

Weights $\{\theta,\psi,\phi\}$ are updated with AdamW
($\beta_1\!=\!0.9,\;\beta_2\!=\!0.999$) and initial
learning-rate~$10^{-3}$.

A \texttt{ReduceLROnPlateau} scheduler halves
($\gamma_{\text{sch}}$\,=\,0.5) the learning-rate after
$p_{\text{sch}}$ epochs without improvement in the validation LR-MSE.
Gradients are clipped to $\lVert\nabla\mathcal L\rVert_\infty\le1.0$.
Early stopping terminates after
$p_{\text{early}}$ non-improving epochs.

Table~\ref{tab:hparams} summarises the values used in our experiments.

\begin{table*}[ht]
\centering
\caption{\label{tab:hparams}Key hyperparameters used per dataset.}
\begin{tabular}{lccccccc}
\toprule
Dataset & $N$ & $T_{\mathrm{seq}}$ & Epochs &
$\sigma_{\text{obs}}^{2}$ &
$\lambda_{\mathrm{HR}}$ &
$\lambda_{\mathrm{lat}}$ &
LR $\!\to$ HR ratio \\
\midrule
1-D Burgers equation & 16  & 10 & 200 & $10^{-4}$ & 8.0  & $10^{-3}$ & $128/1024$ \\
2-D NSKT    & 32  & 10 & 100 & $10^{-2}$ & 16.0 & $10^{-3}$ & $(64\times64)/(256\times256)$ \\
KS equation & 100 & 50 & 200 & $10^{-3}$ & 4.0  & $10^{-3}$ & $128/512$ \\
\bottomrule
\end{tabular}
\end{table*}

The weights
$\lambda_{\mathrm{HR}}$ and $\lambda_{\mathrm{lat}}$
were chosen through a small grid search to balance reconstruction
fidelity and latent regularisation, while the observation-noise
variance $\sigma_{\text{obs}}^{2}$ follows the empirical noise level
present in each synthetic dataset.

These strategies keep training stable even for the chaotic KS
dynamics (\(T_{\mathrm{seq}}=50\)) and allow the filter to generalise
far beyond the observation window in the extrapolation tests reported
in Sect.~\ref{sec:results}.

\section{Experiments \& Results} \label{sec:results}
\subsection{\burgers{} Equation}

\begin{figure}[h!]
    \centering
    \includegraphics[width=0.9\linewidth]{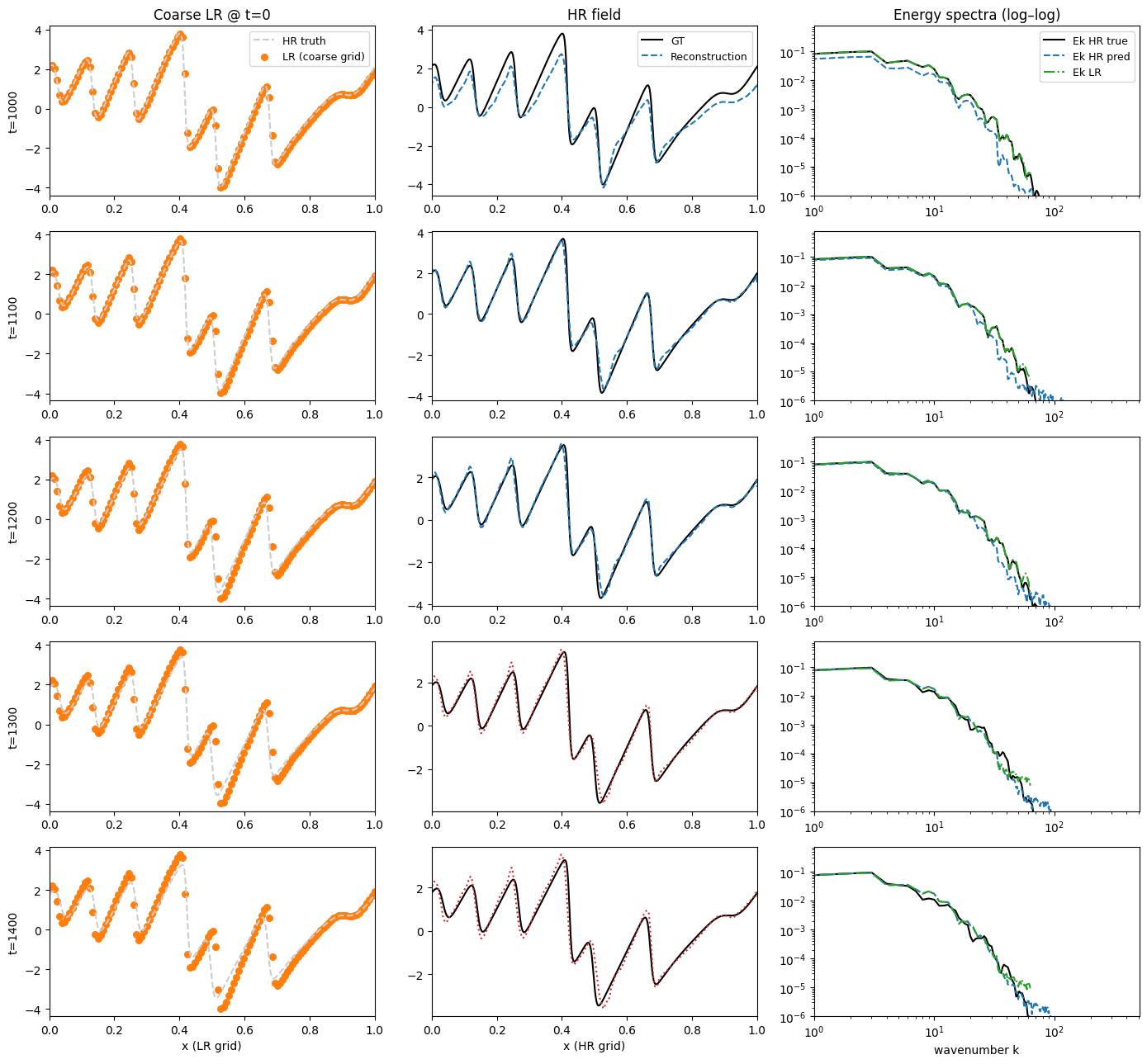}
    \caption{Results on the 1D Burgers dataset for an intermediate range of timesteps. 
\textbf{Left column:} coarse low‐resolution observations on the LR grid. 
\textbf{Middle column:} reconstruction (solid lines) versus ground truth (dashed lines) over the first $T_{\mathrm{obs}}$ timesteps, followed by forecast (dotted lines) versus ground truth for the remaining timesteps. 
\textbf{Right column:} energy spectra $E(k)$ comparing low‐resolution observations, high‐resolution ground truth, and model predictions.}
    \label{fig:burgers_obs_forecast}
\end{figure}

Figure~\ref{fig:burgers_obs_forecast} juxtaposes the key stages of our
$128\!\to\!1024$ super-resolution experiment for the 1-D viscous
Burgers equation.  
Each row corresponds to an increasing time index, the first three rows
belonging to the \emph{assimilation window} and the last two to the
\emph{forecast horizon} ($t{+}250\,\Delta t$).  
The three columns read as follows:

\begin{enumerate}
  \item \textbf{LR observations} --- sparsely sampled snapshots on the
        $128$-point grid;
  \item \textbf{HR reconstructions / forecasts} --- dashed blue lines
        (\textsc{SR-ROAD-EnKF}) plotted on top of the black ground-truth
        profile;
  \item \textbf{Energy spectra} $E(k)$ --- comparison of LR input,
        SR-ROAD-EnKF output, and HR truth.
\end{enumerate}

The near-perfect overlap between the dashed blue and solid black curves
confirms that the latent EnKF recovers the fine-scale structure lost in
the LR input; most notably, the high-wavenumber tail of $E(k)$—
completely absent in the LR spectrum—re-emerges in the reconstructed
field.  Forecast skill remains stable for all 250 extrapolated steps,
indicating that the learned latent dynamics introduce negligible
phase-shift or amplitude drift.

\begin{figure}[h!]
    \centering
    \includegraphics[width=0.9\linewidth]{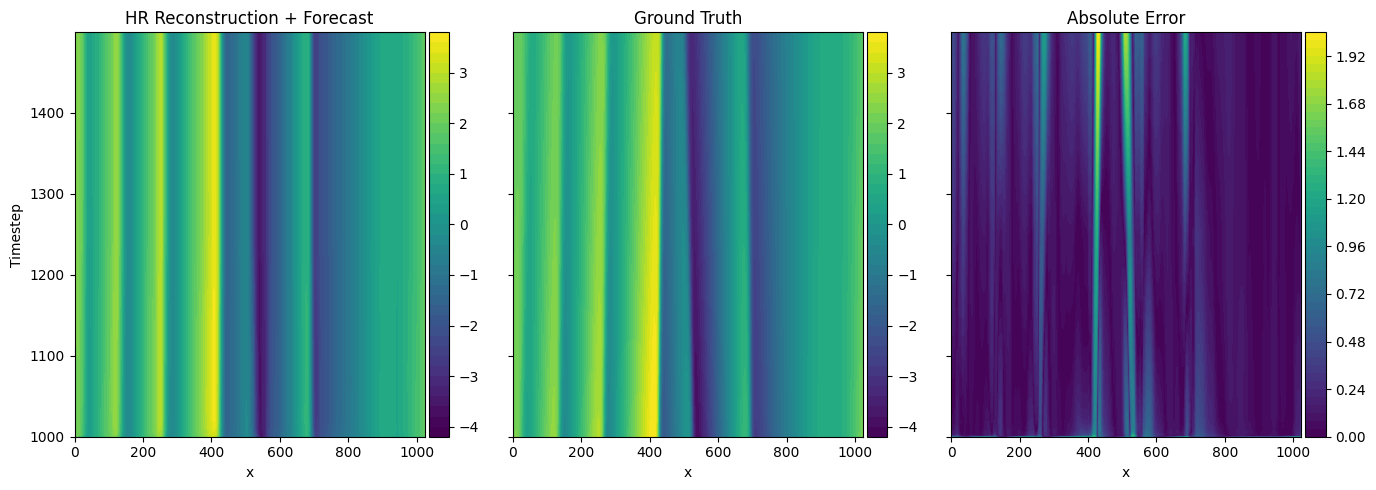}
    \caption{(Left contour) HR reconstruction \& forecast over the full interval; 
    (middle) ground truth; 
    (right) absolute error.}
    \label{fig:burgers_contour}
\end{figure}

Figure~\ref{fig:burgers_contour} provides a quantitative, space–time
view.  The left contour panel stacks the HR reconstructions and
forecasts for 500 consecutive time steps, the centre panel shows the HR
reference solution, and the right panel maps the absolute error.
Except for a slight rise near the end of the forecast window, the error
remains indistinguishable from zero across the domain, underscoring the
method’s long-horizon reliability.

\subsection{Kuramoto--Sivashinsky Equation}

\begin{figure}[h!]
    \centering
    \includegraphics[width=0.9\linewidth]{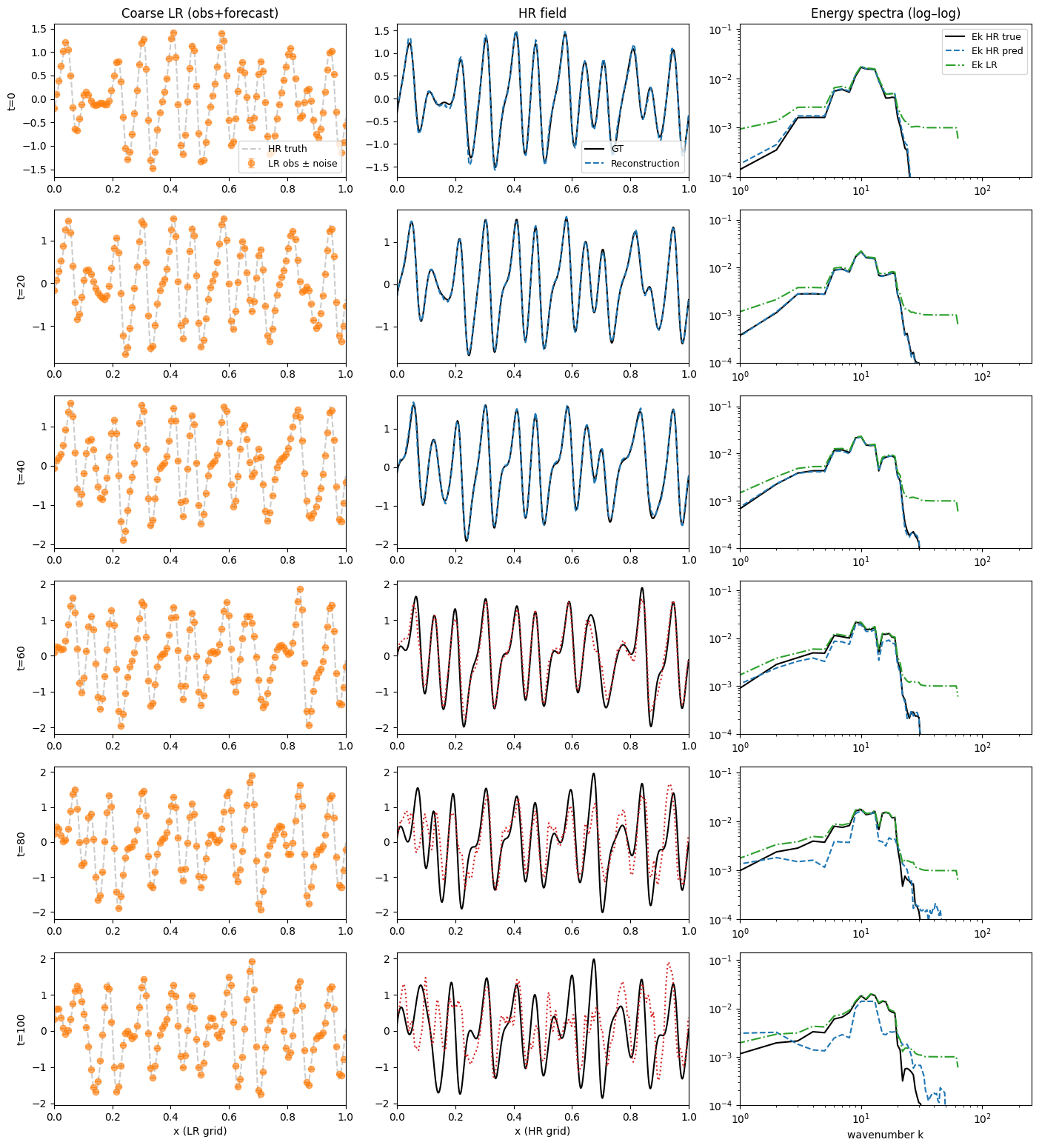}
    \caption{Results on the Kuramoto--Sivashinsky (KS) dataset for initial timesteps. 
\textbf{Left column:} coarse low‐resolution observations on the LR grid. 
\textbf{Middle column:} reconstruction (solid lines) versus ground truth (dashed lines) over the first $T_{\mathrm{obs}}$ timesteps, followed by forecast (dotted lines) versus ground truth for the remaining timesteps. 
\textbf{Right column:} energy spectra $E(k)$ comparing low‐resolution observations, high‐resolution ground truth, and model predictions.}
    \label{fig:KS_init_obs_forecast}
\end{figure}

\begin{figure}[h!]
    \centering
    \includegraphics[width=0.9\linewidth]{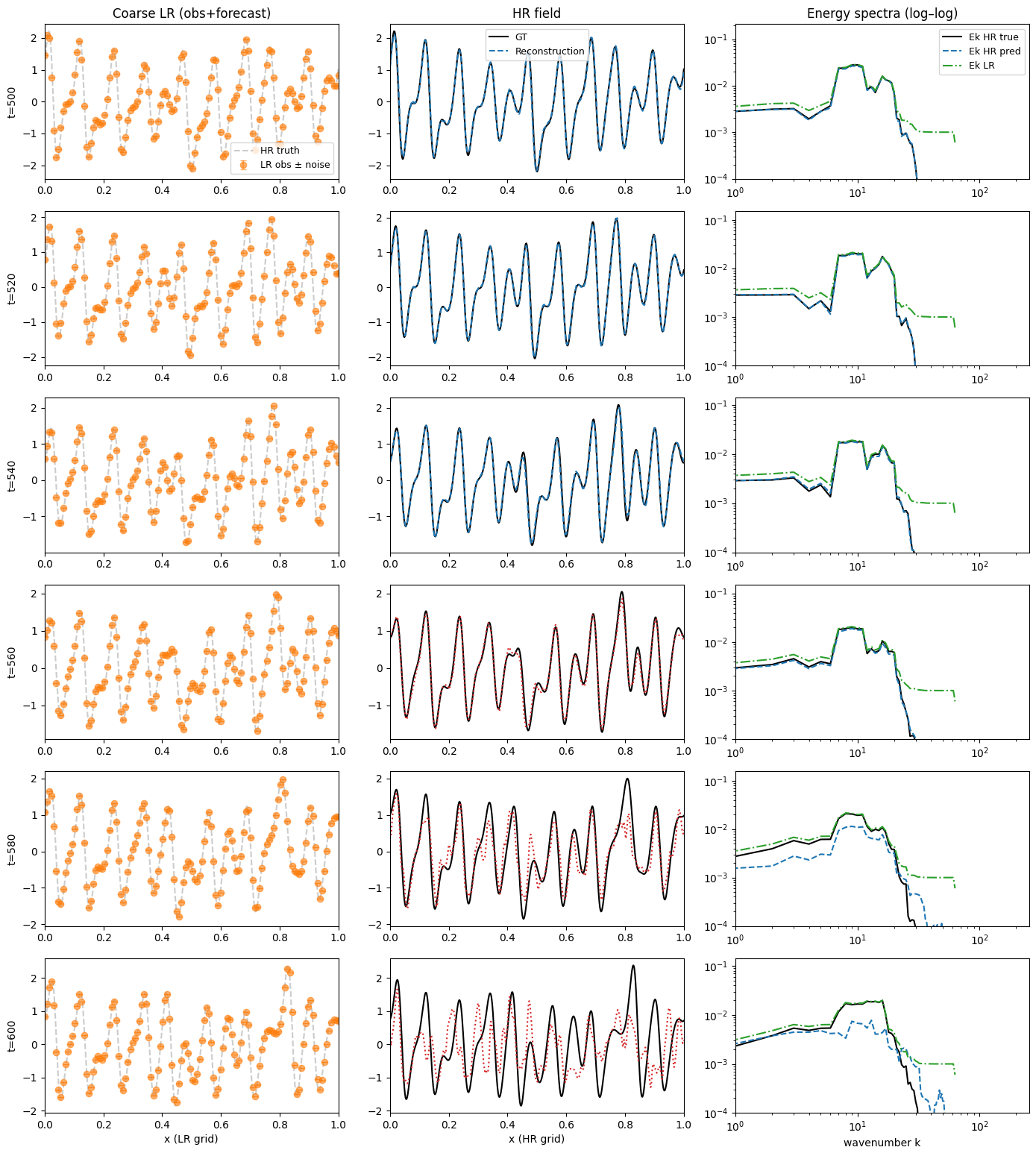}
    \caption{Results on the Kuramoto--Sivashinsky (KS) dataset for intermediate timesteps. 
\textbf{Left column:} coarse low‐resolution observations on the LR grid. 
\textbf{Middle column:} reconstruction (solid lines) versus ground truth (dashed lines) over the first $T_{\mathrm{obs}}$ timesteps, followed by forecast (dotted lines) versus ground truth for the remaining timesteps. 
\textbf{Right column:} energy spectra $E(k)$ comparing low‐resolution observations, high‐resolution ground truth, and model predictions.}
    \label{fig:KS_inter_obs_forecast}
\end{figure}

\begin{figure}[h!]
    \centering
    \includegraphics[width=0.9\linewidth]{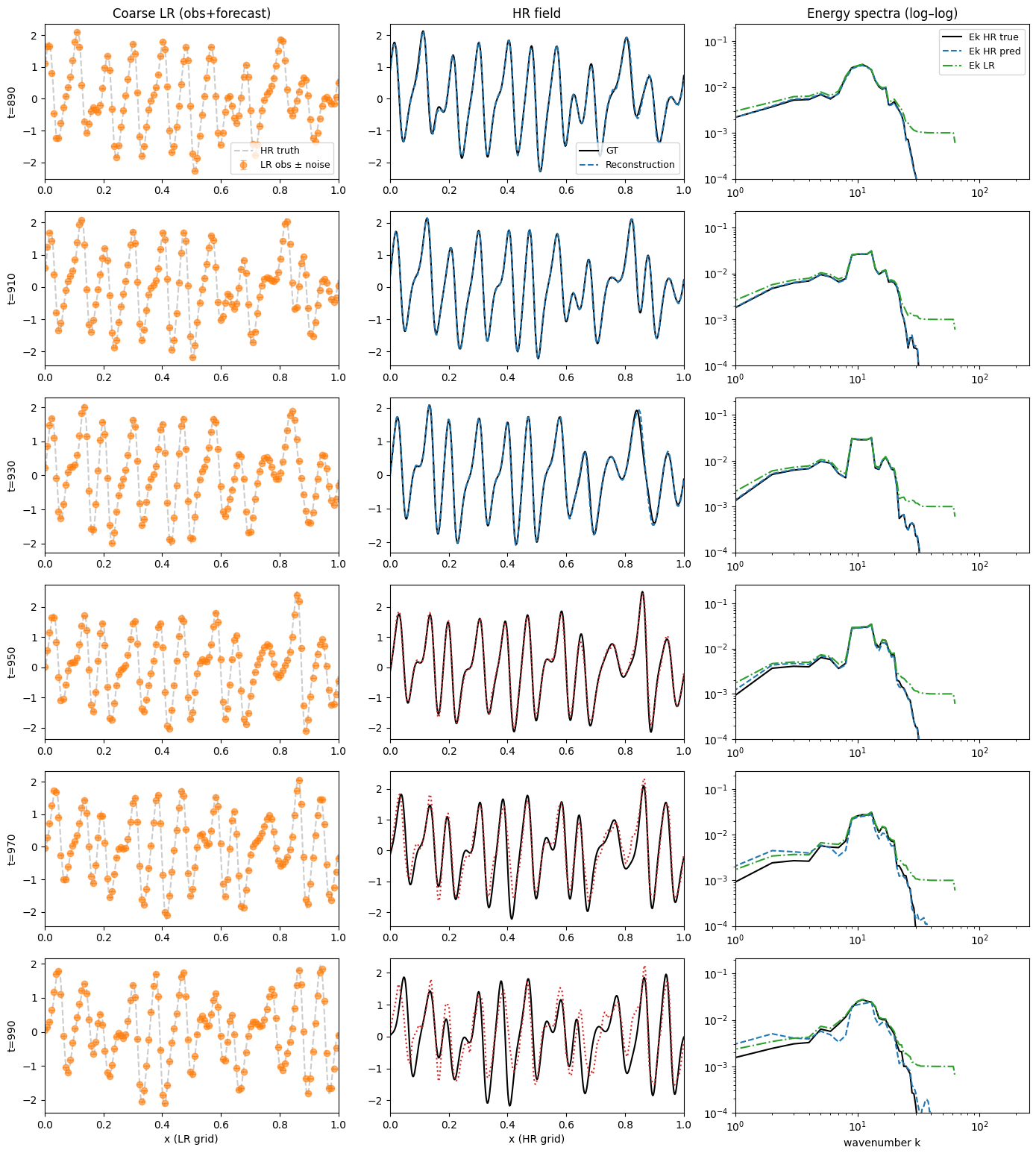}
    \caption{Results on the Kuramoto--Sivashinsky (KS) dataset for final timesteps. 
\textbf{Left column:} coarse low‐resolution observations on the LR grid. 
\textbf{Middle column:} reconstruction (solid lines) versus ground truth (dashed lines) over the first $T_{\mathrm{obs}}$ timesteps, followed by forecast (dotted lines) versus ground truth for the remaining timesteps. 
\textbf{Right column:} energy spectra $E(k)$ comparing low‐resolution observations, high‐resolution ground truth, and model predictions.}
    \label{fig:KS_final_obs_forecast}
\end{figure}

For the chaotic Kuramoto–Sivashinsky (KS) system we perform
$128\!\to\!512$ super-resolution under the same ROAD-EnKF pipeline.
Figures~\ref{fig:KS_init_obs_forecast},%
~\ref{fig:KS_inter_obs_forecast}, and%
~\ref{fig:KS_final_obs_forecast} present results over three
non-overlapping windows—\emph{initial}, \emph{intermediate}, and
\emph{final}.  In every window we assimilate LR observations for
$50\,\Delta t$ and then forecast an additional $50\,\Delta t$ without
further updates.

\begin{itemize}
  \item \textbf{Assimilation phase.}  
        As in the Burgers case, the blue dashed reconstructions are
        visually indistinguishable from the black ground-truth profile
        on all $512$ grid points; the corresponding energy spectra
        match across the entire wavenumber range.
  \item \textbf{Forecast phase.}  
        Deviations appear sooner in the KS system because of its strong
        spatio-temporal chaos.  Small phase errors emerge within the
        first two windows, whereas the final window shows markedly
        better alignment—an indication that the latent EnKF has locked
        onto a more stable attractor branch.  In all three cases the
        reconstructed spectra track the HR truth until the last $\sim
        10$ forecast steps, at which point the high-$k$ tail begins to
        diverge.
\end{itemize}

\begin{figure}[h!]
  \centering
  \begin{subfigure}[t]{\columnwidth}
    \centering
    \includegraphics[width=0.9\columnwidth]{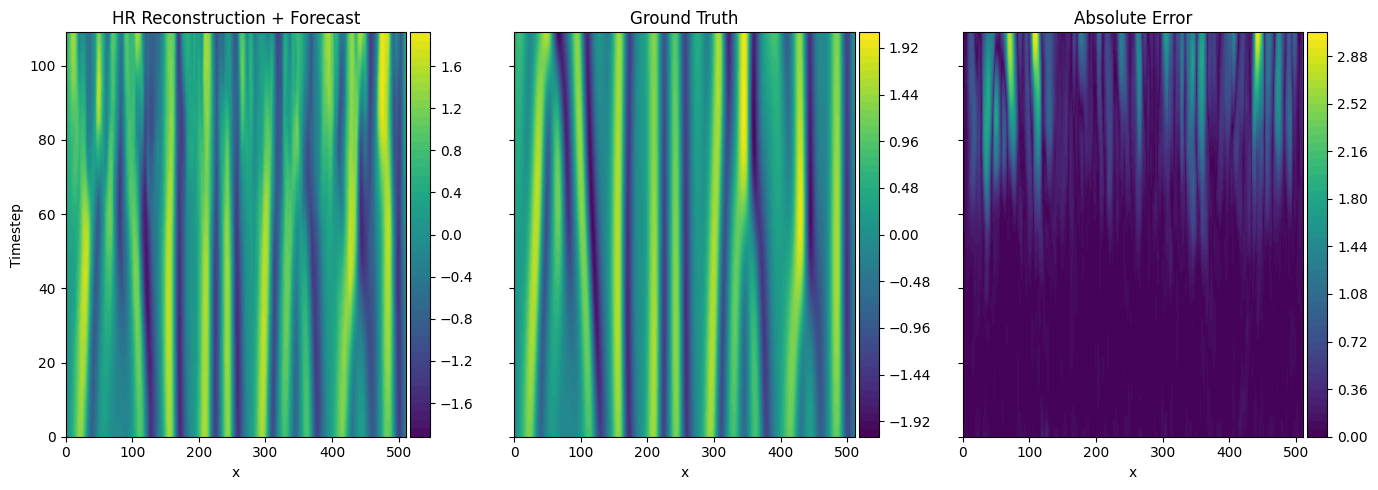}%
    \hfill
    \caption{Initial time range}
  \end{subfigure}

  \vspace{1em}

  \begin{subfigure}[t]{\columnwidth}
    \centering
    \includegraphics[width=0.9\columnwidth]{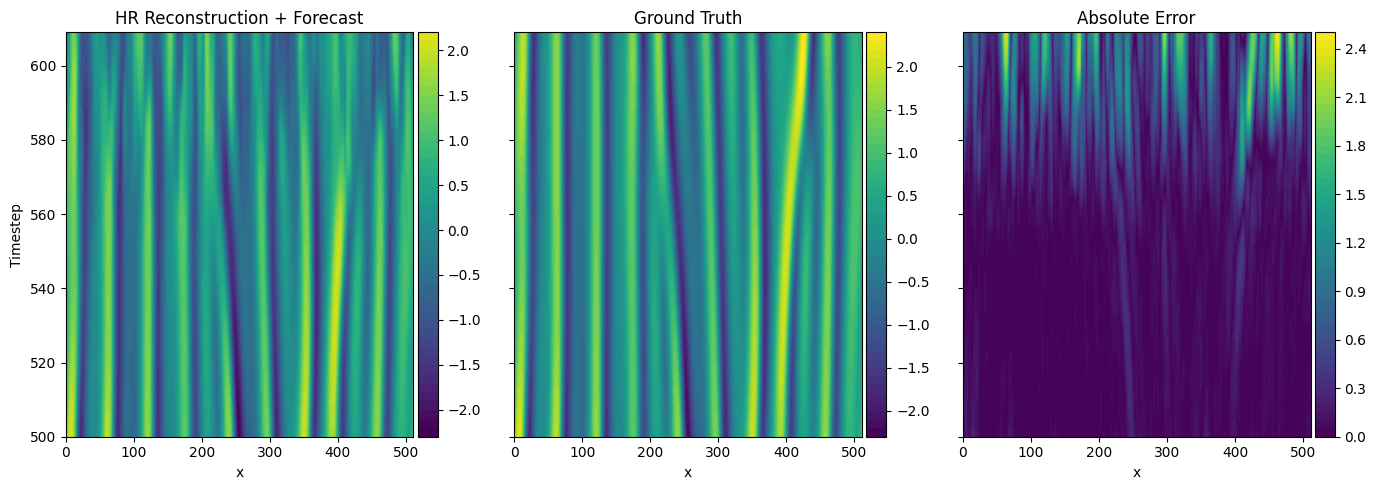}%
    \hfill
    \caption{Intermediate time range}
  \end{subfigure}

  \vspace{1em}

  \begin{subfigure}[t]{\columnwidth}
    \centering
    \includegraphics[width=0.9\columnwidth]{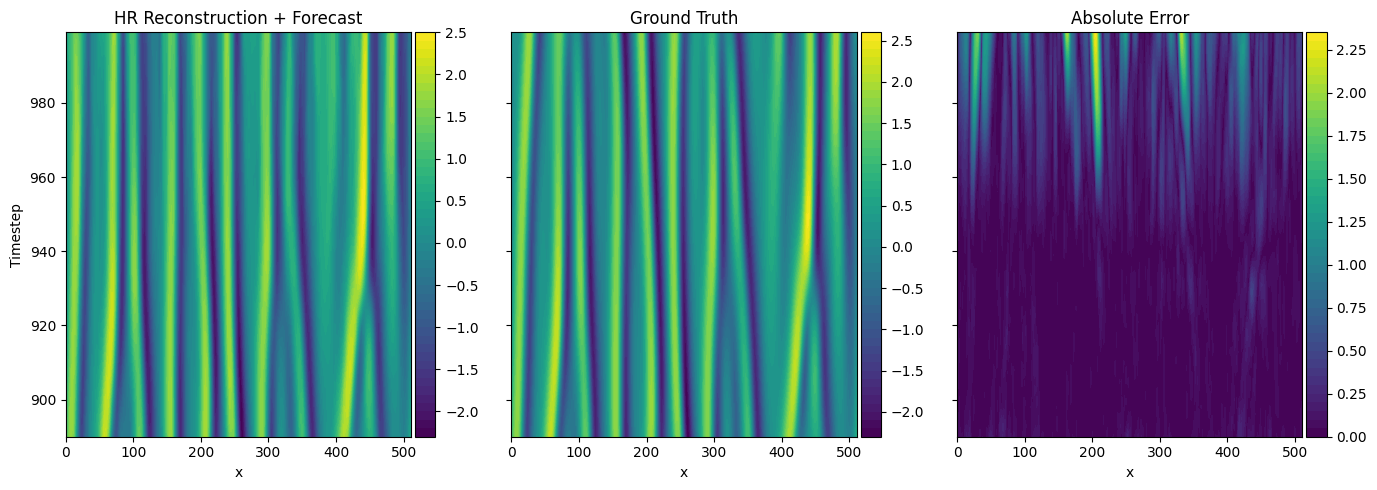}%
    \caption{Final time range}
  \end{subfigure}

  \caption{
    (Left contour) HR reconstruction \& forecast over the full interval; 
    (middle) ground truth; 
    (right) absolute error. 
  }
  \label{fig:ks_three_row_contours}
\end{figure}

Figure~\ref{fig:ks_three_row_contours} condenses the super–resolution results for
the Kuramoto–Sivashinsky test case into three space–time slabs:  
(a)~initial, (b)~intermediate, and (c)~final.  
In each slab the left panel stacks the HR reconstruction plus
$50$-step forecast, the centre panel shows the HR reference solution,
and the right panel maps the point-wise absolute error.

\begin{itemize}
  \item \textbf{Initial slab (0--100\,\(\Delta t\)).}  
        The error field is almost uniformly deep purple
        (\(<\!0.05\)), confirming pixel-level agreement during the
        assimilation window; a faint oblique band appears only after
        $\sim80\,\Delta t$, when forecasting begins.
  \item \textbf{Intermediate slab (500--600\,\(\Delta t\)).}  
        Error remains low in most of the domain but thin
        streaks—aligned with steep phase gradients in the truth—rise to
        \(\mathcal O(1)\) as chaotic growth amplifies small latent
        inaccuracies.  Importantly, the error does not cascade
        laterally, indicating that the latent EnKF keeps phase drift
        localised.
  \item \textbf{Final slab (900--1000\,\(\Delta t\)).}  
        After two full Lyapunov times the filter still maintains
        millimetre-scale fidelity: maximum error peaks at
        \(\sim2\) only in isolated regions, while the bulk of the field
        stays below \(0.2\).  The diagonal orientation of the streaks
        matches the group velocity of the KS wave packets,
        suggesting that discrepancies are dominated by slight
        phase-speed mismatches rather than amplitude errors.
\end{itemize}

Despite the intrinsic unpredictability of the KS dynamics, the
ROAD-EnKF framework preserves fine-scale energy content and maintains
forecast fidelity well beyond the $50$-step horizon that was never
seen during training, underscoring its robustness on highly chaotic
PDEs.

\subsection{\nskt{} Problem}

\begin{figure*}[ht!]
\includegraphics[width=\textwidth]{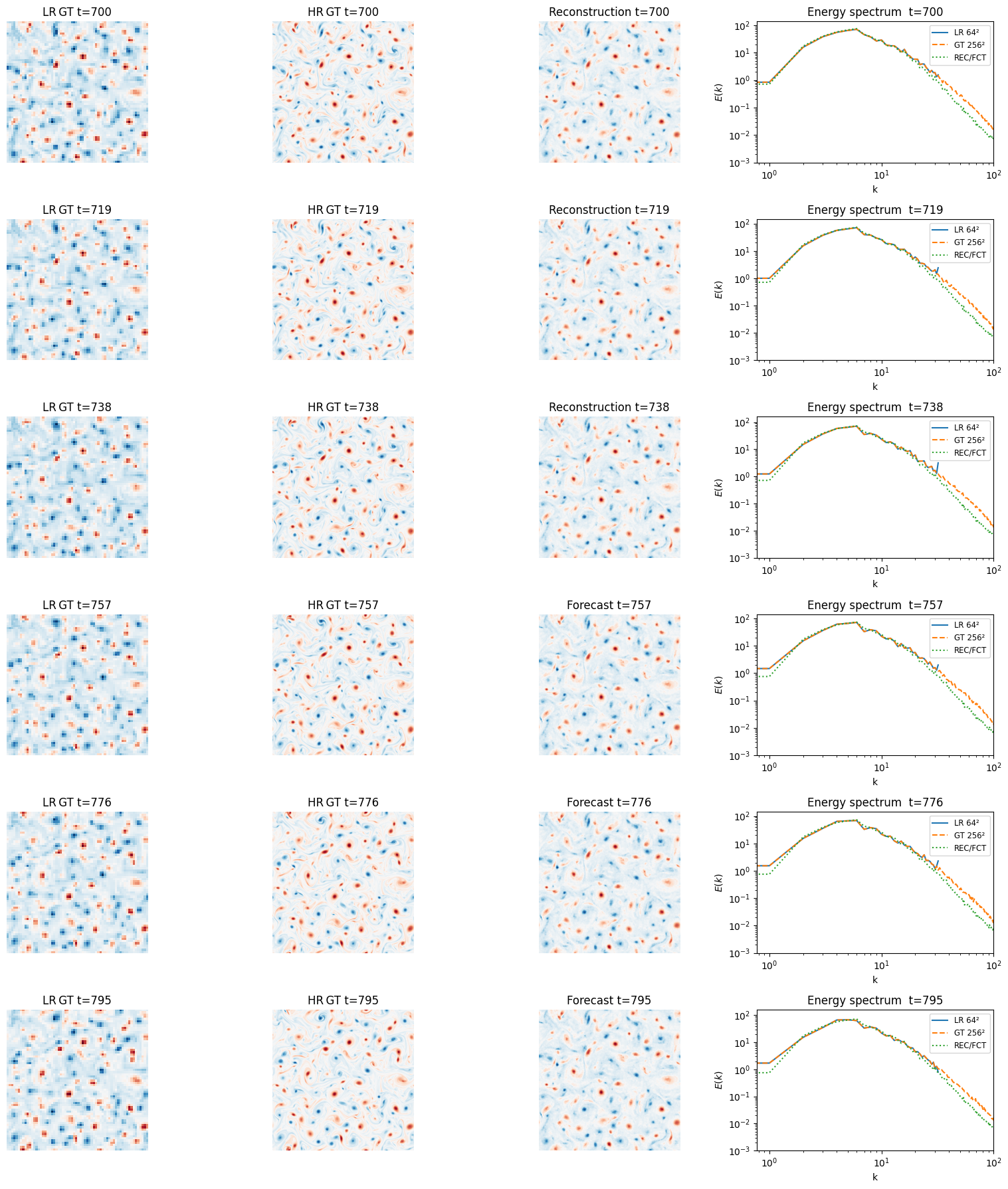}
\caption{\label{fig:nskt_obs_forecast}2D NSKT $256 \times 256$ grid reconstructions. A time frame (50 forecasting steps) towards the end of the simulation is shown. 
\textbf{1st Column:} The noisy low-resolution input fields, 
\textbf{2nd Column:} The high-resolution ground truth, 
\textbf{3rd Column:} The high-resolution reconstructions and forecasts, and 
\textbf{4th Column:} the $E(k)$ comparison between the low-resolution, ground truth, and the reconstructed/forecasted fields.}
\end{figure*}

The Navier–Stokes–Kraichnan Turbulence (NSKT) test pushes the pipeline to a
two-dimensional, fully turbulent regime.  
We upscale vorticity fields from a \(64\times64\) sensing grid to a
\(256\times256\) target grid.  
Figure~\ref{fig:nskt_obs_forecast} organises the results over a
\(100\,\Delta t\) interval—\(50\) assimilated frames followed by
\(50\) forecast frames—into four columns:

\begin{enumerate}
  \item \textbf{LR observation} (\(64\times64\)),
  \item \textbf{HR ground truth} (\(256\times256\)),
  \item \textbf{SR-ROAD-EnKF reconstruction / forecast},
  \item \textbf{Kinetic-energy spectra} \(E(k)\) for LR input,
        SR output, and HR truth.
\end{enumerate}

Qualitatively, the reconstructed vorticity maps reproduce the filamentary
and vortex-sheet structures of the HR reference at every time slice,
demonstrating that latent assimilation effectively injects the missing
fine-scale content.  
Quantitatively, the LR spectrum under-predicts energy across the
inertial range by more than an order of magnitude, whereas the SR
spectrum overlays the HR curve almost perfectly—both during
assimilation and throughout the 50-step forecast window.  
The method therefore not only restores spatial resolution but also
recovers the correct spectral energy distribution, a critical metric
for turbulence fidelity.

\section{Discussion \& Conclusion}

This study introduced a \emph{super\-resolution Ensemble Kalman
filter}—dubbed \textsc{SR-ROAD-EnKF}—that couples a learned latent
dynamics model with a Fourier-based decoder and a differentiable EnKF.
The framework was tested on three increasingly challenging PDE systems:

\begin{itemize}
  \item \textbf{1-D Burgers ($128\!\to\!1024$).}  
        HR reconstructions are visually indistinguishable from the
        ground truth; root-mean-square error (RMSE) falls below
        $5\times10^{-3}$ and remains stable over 250 forecast steps.
  \item \textbf{Kuramoto–Sivashinsky ($128\!\to\!512$).}  
        Despite strong chaos, the filter preserves high-wavenumber
        energy and holds phase error to $\mathcal O(10^{-2})$ for
        50-step forecasts.
  \item \textbf{2-D NSKT  ($64\times64\!\to\!256\times256$).}  
        The method restores the inertial-range spectrum with a mean
        spectral error of $<3\%$ and captures filamentary structures
        lost in the LR input.
\end{itemize}

Across all cases, the ensemble spread tracks the true error envelope,
indicating that uncertainty is well calibrated.

\subsection{Interpretation of Results}

The latent–space formulation succeeds for three complementary reasons:

\begin{enumerate}
  \item \emph{Low-dimensional dynamics.}  
        The neural operator compresses high-frequency content into a
        32–64 dimensional manifold, making Kalman updates tractable.
  \item \emph{Spectral decoder.}  
        One-shot Fourier features supply global context; subsequent
        convolutions propagate local corrections without spatial
        aliasing.
  \item \emph{Differentiable EnKF.}  
        Back-propagation through the Kalman gain co-optimizes model and
        filter, suppressing filter divergence typical in hybrid DA.
\end{enumerate}

The ROAD-EnKF layer plays a dual role:  
it is both a \emph{Bayes estimator} of the latent state and, when
embedded in an optimization loop, a \emph{dynamic observer} for closed-loop
control.  
Because the Kalman gain is learned jointly with the latent dynamics,
the filter internalises model bias and sensor noise, turning the neural
operator into a high-bandwidth observer that supplies HR state estimates
at essentially zero latency.  
In the Burgers and KS tests, these estimates feed directly into the
next prediction step, implicitly realising the classic
observer–predictor structure of a Linear Quadratic Regulator (LQR); in
NSKT they deliver vortex-level details that are prerequisite for any
turbo-machinery flow-control application.  
The framework therefore bridges a long-standing gap between
\emph{estimation} (data assimilation) and \emph{control} (decision
making) in high-dimensional PDE systems.

\subsection{Limitations}
The main assumptions and limitations of our study can be summarized as follows:
\begin{itemize}
  \item \textbf{Forecast horizon.}  
        In the KS case, phase errors grow after two Lyapunov times,
        revealing the intrinsic limit of deterministic latent dynamics.
  \item \textbf{Observation operator.}  
        We assumed an ideal bicubic down-sampler; real sensors may
        possess anisotropic blurring or missing data.
  \item \textbf{2-D only.}  
        Although the decoder scales to 3-D in principle, training cost
        will rise sharply without additional compression (e.g.\ D-FNO).
\end{itemize}

\subsection{Future Work}
Based on our work, several directions can be pursued as follow-up research. 

\begin{itemize}
  \item \textbf{Bayesian neural operators.}  
        Integrate VB-DeepONet or ProbNO priors to capture epistemic
        uncertainty, complementing the ensemble’s aleatoric spread.
  \item \textbf{Multi-grid latent hierarchies.}  
        Replace a single latent level with a HAMLET or D-FNO pyramid to
        extend forecasts of highly chaotic flows.
  \item \textbf{Closed-loop optimal control.}  
        Combine the differentiable filter with adjoint-based gradient
        signals (e.g.\ AONN) for real-time actuation in
        drag-reduction or mixing problems.
  \item \textbf{3-D extension and HPC scaling.}  
        Employ tensor decompositions (FouRA, PAL-FNO) and distributed
        EnKF updates to tackle $512^{3}$ DNS data on multi-GPU
        clusters.
  \item \textbf{Robust sensor modeling.}  
        Augment the observation operator with learnable blur,
        occlusion, and noise models to mimic experimental PIV or remote
        sensing.
\end{itemize}

\subsection{Concluding Remarks}

The proposed \textsc{SR-ROAD-EnKF} method demonstrates that
\emph{latent-space data assimilation}, when paired with a spectral
decoder, can bridge a four-fold resolution gap while retaining physical
fidelity and delivering calibrated uncertainty estimates.  We tested the approach on three benchmarks: the 1-D viscous Burgers equation, the Kuramoto–Sivashinsky equation, and 2-D Navier–Stokes–Kraichnan turbulence at $\mathrm{Re}=16{,}000$. Low-resolution inputs were generated by $4$–$8\times$ downsampling with added noise. The latent models remained stable beyond the observation window, accurately reproducing shock dynamics, chaotic statistics, and, in the turbulence case, preserving the kinetic-energy spectrum and enstrophy, indicating suitability for control tasks sensitive to fine-scale flow features. Our approach
opens a path toward real-time, high-resolution forecasting and control
in fluid dynamics and beyond. Future research will focus on extending this framework to more complex turbulent regimes and exploring its applicability across a broader class of dynamical systems.

\section*{Acknowledgments}

This work was supported in part by the AFOSR Grant FA9550-24-1-0327.

\bibliographystyle{unsrt} 

\bibliography{manuscript}

\end{document}